\documentclass[conference,10pt]{IEEEtran}

\usepackage{amssymb}
\usepackage{amsmath}
\usepackage{amsthm}
\usepackage{mathtools}
\usepackage{amsfonts}
\usepackage{txfonts}
\usepackage{dsfont}
\usepackage{times}
\usepackage{siunitx} 
\usepackage{soul} 
\usepackage{bm}
\usepackage{pifont}
\usepackage{listings}       
\usepackage{stackengine}

\def\dual#1{\expandafter\dual@aux#1\@nil}   
\def\dual@aux#1/#2\@nil{\begin{tabular}{@{}c@{}}#1\\#2\end{tabular}}

\usepackage[T1]{fontenc}
\usepackage[latin1]{inputenc}
\usepackage{authblk}
\usepackage{color}
\usepackage{verbatim}
\usepackage{alltt}
\usepackage[ruled,vlined,lined,linesnumbered]{algorithm2e}
\usepackage[noend]{algpseudocode}
\usepackage{ragged2e}
\usepackage[font=sf,bf]{caption}
\usepackage{comment}
\usepackage{hyperref}
\usepackage[final]{pdfpages} 
\usepackage{pdflscape} 
\usepackage{enumerate}
\usepackage[noadjust]{cite}
\usepackage{enumitem}
\usepackage[final]{pdfpages} 
\usepackage{balance}
\usepackage{fancyhdr, lipsum}
\usepackage{textcomp}
\usepackage[nottoc]{tocbibind}
\usepackage{tcolorbox}
\usepackage{xcolor,colortbl}
\usepackage{setspace}

\SetAlFnt{\small}
\SetAlCapFnt{\small}
\SetAlCapNameFnt{\small}
\SetAlCapHSkip{0pt}

\usepackage{epsfig}
\usepackage{epstopdf}
\usepackage{float}
\usepackage{graphics}
\usepackage{graphicx}
\usepackage{psfrag}
\usepackage[position=bottom]{subfig}
\usepackage[section]{placeins}
\usepackage[export]{adjustbox}
\usepackage{wrapfig}
\usepackage{rotating}

\usepackage{longtable}
\usepackage{multirow}
\usepackage{booktabs}
\usepackage{array}

\usepackage{todonotes}
\usepackage{url}

\newtheoremstyle{mytheoremstyle}{2pt}{1pt}{\itshape}{}{\bfseries}{.}{.5em}{} 
\theoremstyle{mytheoremstyle}
\usepackage{xpatch}
\makeatletter
\xpatchcmd{\proof}{\topsep6\p@\@plus6\p@\relax}{}{}{}
\makeatother

\newtheorem{definition}{Definition}

\newtheorem{coverage}{Coverage Artefact}

\newcommand{\at}[2][]{#1|_{#2}} 

\SetCommentSty{commentFont}

\setlongtables
\makeatletter

\title{Methodology for Biasing Random Simulation for Rapid Coverage of Corner Cases in AMS Designs}

\author[1]{Sayandeep Sanyal}
\author[2]{Ayan Chakraborty}
\author[1]{Pallab Dasgupta}
\author[1]{Aritra Hazra}
\affil[1]{Dept. of Computer Science \& Engineering}
\affil[2]{Dept. of Electronics \& Electrical Communication Engineering}
\affil[ ]{Indian Institute of Technology Kharagpur}
\date{30 April 2021}

\begin{document}

	\thispagestyle{empty}
	\maketitle
	
\begin{abstract}
	Exploring the limits of an Analog and Mixed Signal (AMS) circuit by driving appropriate inputs has been a serious challenge to the industry. Doing an exhaustive search of the entire input state space is a time-consuming exercise and the returns to efforts ratio is quite low. In order to meet time-to-market requirements, often suboptimal coverage results of an integrated circuit (IC) are leveraged. Additionally, no standards have been defined which can be used to identify a target in the continuous state space of analog domain such that the searching algorithm can be guided with some heuristics. In this report, we elaborate on two approaches for tackling this challenge -- one is based on frequency domain analysis of the circuit, while the other applies the concept of Bayesian optimization. We have also presented our results by applying the two approaches on an industrial LDO and a few AMS benchmark circuits.
\end{abstract}

\section{Introduction}
\label{sec: intro}

    Circuit validation has always taken up the lion's share of time (and manpower) in the life cycle of circuit design and fabrication. Most of the circuit design community follow a bottom-up designing approach. The IPs are designed and verified in isolation before integrating them with other IPs and re-doing the verification steps. At IP level verification phase, the design complexity is low, hence simulations are faster. This enables the verification engineer to run a good number of tests on the IP and do rigorous testing. Accordingly, a good amount of bugs are identified and fixed at this stage. On proceeding to higher levels of hierarchies, the time taken to simulate the design increases hence the feasibility of running a large amount of simulations comes down. Due to this several critical bugs may escape at these stages. If the design bug is caught at a latter stage of system integration it becomes more painful a task to descend to the IP level and fix it. Moreover, it invalidates a portion of verification results that have been collected so far, and hence these steps have to be repeated before the final sign-off of the design. The situation becomes worse if the bug gets activated during the field run of the chip. Since many of these electronic chips are finding their application in safety-critical domains like automotive and health monitoring systems, a buggy operation at this level could be fatal. Hence, it is of utmost importance to catch these bugs at an early stage.
    
    Since analog designs deal with continuous state space validating the design integrity for all possible input scenario is theoretically non-trivial and practically infeasible. Designers and verification engineers rely on random (or constraint-random) simulation environments that carries out multiple simulations by randomly choosing input parameters from a given state space. One such widely practiced technique is MonteCarlo simulation. A drawback in these techniques is that they randomly choose the input vectors from the given input space, out of which many data points may not lead to disclosing new output response of the circuit. Thus, the time and computational resources that go behind these simulations do not bear any fruit. Whereas a guided random simulation environment which is able to do a surgical cut of the input space will be much appreciated as it prunes down the search space and proceeds towards the goal state in fewer iterations.

    We believe that our work on analog coverage \cite{covert, cross_coverage} can be utilised effectively in this context leading to comprehensive coverage analysis for AMS designs. Analog coverage deals with quantifying the input and output state space of a design. We have designed and developed the framework for analog coverage based on the notion of enumerable regions in the state space called {\em bins}. This facilitates us to assign a target or goal state, in the form of {\em bin(s)}, to the random sampler and biasing it such that it reaches the goal state in fewer iterations. This can be beneficial in the following aspects. 
    
    \begin{enumerate}
        \item Portions of the target space can be tagged as visited and the sampling engine can focus more on visiting the uncharted territories. This improves the quality of verification, often referred to as coverage, by testing the design in regions of operation which have not been encountered before.
        
        \item It facilitates the search for any reachable bad state in the design. A bad state corresponds to the state where the response of the system is not adhering to this specification. If no such state is found during exploration it bestows confidence on the proper functioning of the design.
    \end{enumerate}

	\begin{figure*}
		\centering
		\subfloat[Range Coverage]{
			\includegraphics[width=0.4\textwidth]{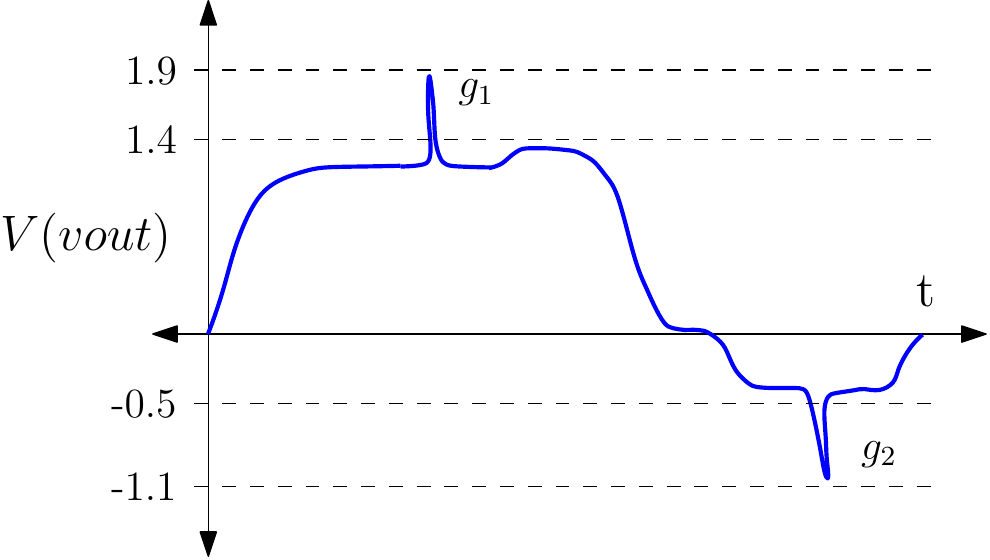}
			\label{fig : deglitch}
		}\qquad
		\subfloat[Level Coverage]{
			\includegraphics[width=0.4\textwidth]{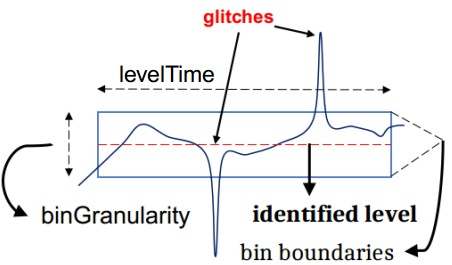}
			\label{fig : level}
		}
		\caption{Some AMS Coverage Primitives}
		\vspace{-0.4cm}
	\end{figure*}
    
    In this report, we present two such techniques for biasing the random simulation for facilitating the exploration of unvisited or uncovered states/regions of the design. The first approach is based on the frequency domain analysis of the design, while the second approach applies the technique of Bayesian optimization for scaling the problem. The rest of the report is organised as follows. Section \ref{sec: coverage_types} presents the functional artefacts that we have developed specifically for analog coverage. Section \ref{sec: coverage overview} gives an overview of the overall coverage problem. Section \ref{sec: problem formulation} details out the two approaches along with their results on some industrial and benchmark circuits. Section \ref{sec: conclude} holds our concluding remarks.

	\section{AMS Coverage Artefacts}
	\label{sec: coverage_types}

	As opposed to enumerable domains of digital coverage bins, the notion of intervals are at
	the heart of analog coverage bins.
	
	\begin{definition} \label{def: bin}
		{\bf [ Bin ]}\\
		A {\em bin}, $\beta$, can be defined as a nonempty convex subset of $\mathbb{R}$ expressed as $ [a:b],~ 
		(a:c),~ [a,c)$ and $ (a,c]$ where $a,b,c \in \mathbb{R}$ and $b \geq a, c>a$. Here, $a$, $b$ and $c$ are 
		called {\em bin boundaries}; $l(\beta)$ and $r(\beta)$ are used to denote the left and right boundaries 
		of bin $\beta$, respectively. \qed
	\end{definition}

	The input to the coverage monitor is time-stamped data of the signals for which the 
	coverage is to be computed. Formally:
	
	\begin{definition}
		\label{def: input}
		{\bf [ Input Function ]}\\ 
		Input function, $\mu$, is defined as a mapping, $\mu (\varv,t): \mathbb{R}^{+} \rightarrow \mathbb{R}$ 
		for a signal $\varv$ and time $t$. Similarly, for a system having the set of signals as, 
		$\overrightarrow{\varv} = \{ v_1, v_2, \dots, v_k\} $, the mapping is defined as, $ \mu 
		(\overrightarrow{\varv},t): \mathbb{R}^+ \rightarrow \mathbb{R}^k$. \qed
		
	\end{definition}
	
	\begin{definition}
		\label{def: sim_trace}
		{\bf [ Simulation Trace ]}\\
		A simulation trace, $\tau$, is a mapping $\tau: \mathbb{R}_{\geq 0} \rightarrow \mathbb{R}^{|V|}$, where 
		$V = \{v_1,~v_2,~...,~v_n\}$ is the set of variables (Boolean and Real) representing signals of the 
		system and $|V|$ indicates the cardinality of the variable set. \qed
		
	\end{definition}
	
	The formal definition of each AMS Coverage artefact follows.

	\begin{coverage}{\bf [Range Coverage]}
		\label{subsubsec: range}
		
		Range coverage is aimed at reporting the range that a signal has observed throughout the entire simulation 
		run. 	For a signal $\varv$, the range coverage returns an interval $[a,b]$ ($a,b \in \mathbb{R}$ and $a 
		\leq b$) such that $\forall x$, $x \in {\tt range}\big{(}\mu (v,t)\big{)}$, when $a\leq x$ and $b \geq x$. 
		
	\end{coverage}
	
	\begin{coverage}{\bf [De-glitched Range Coverage]}
		\label{subsubsec: deglitch}
		
		For a given value of {\em deglitchingTime}, $\delta_t$, all overshoots or undershoots in the signal whose width is less than $\delta_t$ is regarded as a glitch. De-glitched range coverage type will compute the range of the signal when these glitches in it, if any, are discarded.	
	\end{coverage}
	
	As shown in Figure \ref{fig : deglitch} a range coverage 
	artefact on the signal {\tt V(vout)}, the output voltage of an LDO, will return the interval $[-1.1 : 1.9]$. 
	Whereas a de-glitched range coverage will identify the spikes ($g_1$) and the trough ($g_2$) as glitches 
	(depending on the given de-glitch time) and ignore them. Thus reporting the de-glitched range as $[-0.5 : 
	1.4]$.
	
	\begin{coverage}{\bf [Level Coverage]}
		\label{subsubsec: level}
		
		The discrete levels where a signal has settled at for a time being, will be captured through level 
		coverage. Given a value to {\em levelTime}, $\delta_l$, the signal has to settle with in a bin for a minimum of time duration of $\delta_l$ for the bin to be treated as a {\em level}.
	\end{coverage}
	
	Figure \ref{fig : level} shows the necessary conditions for a signal to be associated with a discrete level 
	value. As shown, the de-glitched signal has to stay within a bin of size {\em k} ({\tt binGranularity}) for 
	at least $\delta_l$ ({\tt levelTime}) amount of time such that the mean of that bin boundary is identified 
	as a level of the signal.

	\begin{coverage}{\bf [{\em ddt} Coverage]}

		\label{subsubsec: ddt}
		
		The intent of a {\em ddt} coverage will be the report how steeply the signal rises or falls. The slope of 
		signal $\varv$ is computed at periodic time points separated by the given value of {\em timeGranularity} 
		$\delta_m$. Slope at any such time point $\hat{t}$ is computed as follows.
		$$ \mu '(\varv,\hat{t}) =  \frac{d\varv}{dt} \at[\Bigg]{t = \hat{t}} = \frac{\mu (\varv, \hat{t} + \delta_m) - \mu (\varv, \hat{t})}{\delta_m} $$

		{\em ddt} coverage of signal $\varv$ returns the interval $[a:b]$ ($a,b \in \mathbb{R}$ and $a\leq b$), such 
		that $\forall \hat{t}$, $\mu'(\varv, \hat{t}) \geq a$ and $\mu'(\varv, \hat{t}) \leq b$.
	\end{coverage}

	\begin{coverage}{\bf [Delay Coverage]}
		\label{coverage : delay}
		
		Delay coverage computes the range of delays between an ordered pair of events. The delay between successive occurrences of events $E_1$ and $E_2$ at times $t_{E_1}$ and $t_{E_2}$ respectively is captured by $delay(E_1, E_2) = t_{E_2} - t_{E_1}$. The delay coverage range is:
		$$ \mathcal{B}_{delay}^{E_1, E_2} = [min(delay(E_1,E_2)) : max(delay(E_1,E_2))] $$
		It may be noted that the delays are discrete, and hence this range is not continuous. 
	\end{coverage}

	\begin{figure*}
		\centering
		\includegraphics[width=\textwidth]{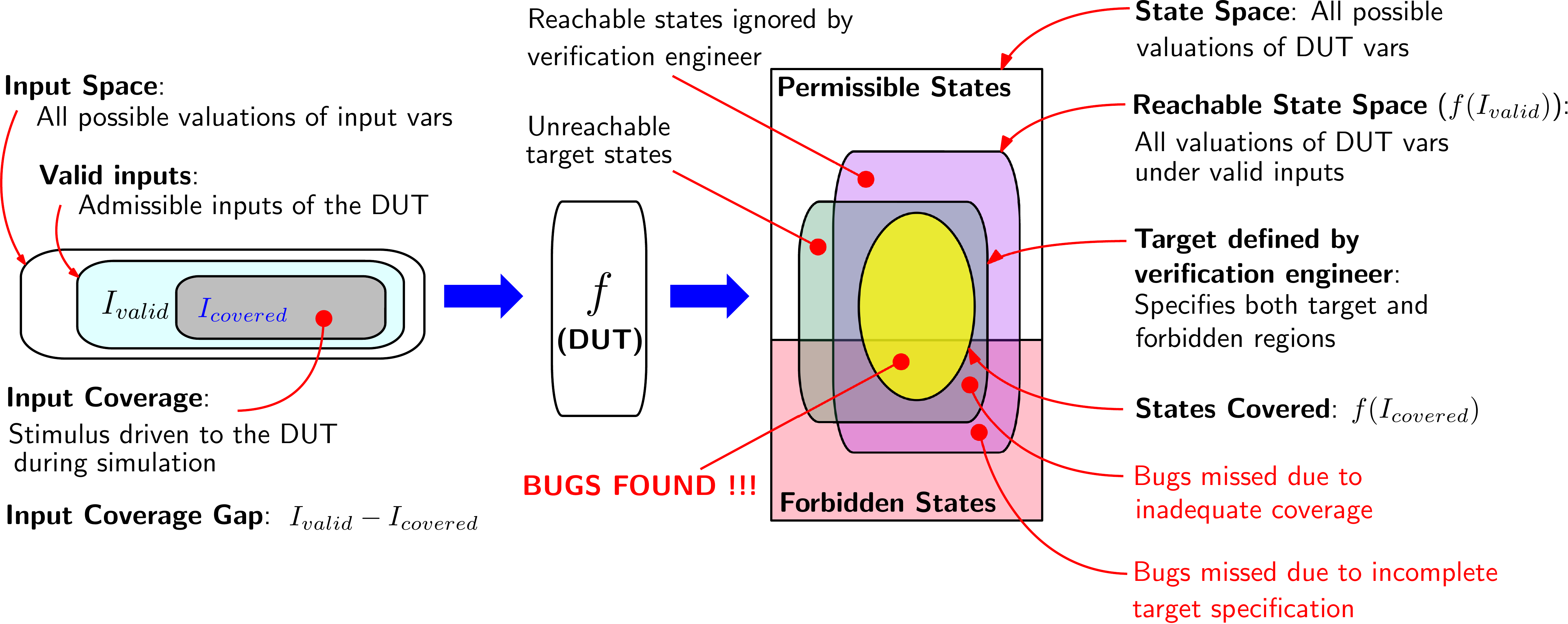}
		\caption{Coverage Perspective}
		\label{fig: coverage overview}
	\end{figure*}

	\begin{coverage}{\bf [Frequency Coverage]}
		\label{subsubsec : frequency}
		
		Given a real value $ \lambda $ and a signal $ \varv $, the frequency of $ \varv $ about $ \lambda $ is defined as the number of crossings of $ \lambda $ by $\varv$ per unit time (or the number of zero crossings of $\varv - \lambda$). The frequency coverage of $\varv$ is the range: 
		$$ \mathcal{B}_{freq}^{\varv, \lambda} = [f_{min} : f_{max}] $$ 
		captured over various time windows. It may be noted that the frequencies are discrete, and hence this range is not continuous.
	\end{coverage}
	
	With the formalism of these AMS coverage artefacts in place, we know proceed into the details on how we structure the analog coverage state space around these artefacts.

	\section{Coverage Overview}
	\label{sec: coverage overview}

	The input space and state space of analog components are infinite and dense. Hence, analog coverpoints cannot be enumerated as in digital. Consequently, specification and analysis of coverage in the analog context is based on {\em regions} defined by {\em predicates over real variables} (PORVs). Fig.~\ref{fig: coverage overview} provides a perspective on coverage analysis. Consider an analog design, $f$, having the set of input variables, $\Sigma$, and the set of internal state variables, $S$. Then:
\begin{itemize}
    \item The {\em input space}, ${\cal I}$, consists of all possible valuations of the variables in $\Sigma$.
    \item The {\em admissible input space}, ${\cal I}_{valid} \subseteq {\cal I}$, defines the domain of $f$, namely the input valuations under which $f$ is supposed to function.
    \item The {\em covered input space}, ${\cal I}_{covered} \subseteq {\cal I}_{valid}$, denotes the input valuations that have been covered, that is, the verification subjected the design to these input valuations.
    \item The {\em input coverage gap} is ${\cal I}_{valid} \setminus {\cal I}_{covered}$.
    \item The {\em state space} of $f$, consists of all possible valuations of the variables in $S$. The design specification partitions the state space into the {\em permissible states}, $S_{valid}$ and the {\em forbidden states}, $S_{invalid}$. The design specification entails that the forbidden states must not be reached.  
    \item The {\em reachable state space}, denoted as $f({\cal I}_{valid})$, is the set of states reachable under admissible inputs. Typically the exact contours of the reachable state space is unknown to the design engineer, and the aim of verification is to determine whether $f({\cal I}_{valid}) \cap S_{invalid} = \emptyset$.
    \item The {\em target state space}, ${\cal T}$, is defined by the coverage target specifications in CoveRT's specification language. The target is divided into two parts namely:
    \begin{itemize}
        \item ${\cal T}_{valid} = T \cap S_{valid}$, which denotes the space which the design is expected to visit, and
        \item ${\cal T}_{invalid} = T \cap S_{invalid}$, which represents the forbidden regions. Covering any bin in ${\cal T}_{invalid}$ is indicative of a design bug. 
    \end{itemize}
    Since the target is specified based on the designer's understanding, it is possible that some states in the target region are not actually reachable by the design, and therefore 100\% coverage of the target is not achievable. It is also possible that states outside the target region are actually reachable, and some of them belong to $S_{invalid}$, that is, forbidden states.
    \item The {\em covered state space}, denoted as $f({\cal I}_{covered})$ represents the design states covered during verification. Then:
    \begin{itemize}
        \item CoveRT reports bugs when $f({\cal I}_{covered}) \cap T_{invalid} \neq \emptyset$.
        \item The entire bug space is $f({\cal I}_{valid}) \cap S_{invalid}$. Bugs residing in $(f({\cal I}_{valid}) \cap S_{invalid}) \setminus f({\cal I}_{covered})$ are missed during simulation.
    \end{itemize}
    \item The {\em coverage gap} is ${\cal T}_{valid} \setminus f({\cal I}_{covered})$
\end{itemize}
A design is typically subjected to many different tests. Each test contributes to coverage, that is, each test adds to ${\cal I}_{covered}$. The regions (bins) covered by each test is maintained by CoveRT in a database. This allows CoveRT to compute the accumulated coverage in $f({\cal I}_{covered})$ and report the remaining coverage gap with respect to the specified target.

\section{Covering Corner Cases in AMS Designs}
\label{sec: problem formulation}

With the AMS coverage artefacts and a simulation-based tool, CoveRT \cite{covert}, to compute these, it does not suffice to report the computed coverage values. One of the primary intentions of coverage based verification methodology is to drive the design under test (DUT) into regions of operation that it has not seen before. Unless done so, the manifestation of certain bugs can be missed during the verification phase. Speaking in terms of coverage, which points from the region $I_{valid} \setminus I_{covered}$ are to be chosen such that the DUT is driven into the unexplored territory of $f(I_{valid}) \setminus f(I_{covered})$, that is, closing the coverage gap.

Currently this is achieved by doing a MonteCarlo simulation of the DUT by choosing the input parameters from $I_{valid}$. One major drawback is that MonteCarlo simulations are require a large number of simulations to be run which is computationally expensive. Even though such large number of simulations may be possible at a lower design hierarchy, IP or block level, it becomes infeasible to perform at a higher hierarchy like, sub-system or SOC level. Hence the requirement for intelligently guiding/biasing the MonteCarlo simulations such that uncovered regions are visited in a reduced number of simulations. We have designed two such approaches mentioned as follows.
\begin{enumerate}
    \item A frequency domain analysis based approach where we analyse the bode plot of the DUT to find an appropriate input signal which lead to spread the current boundary of $f(I_{covered})$.
    \item A Baysian optimization based approach where we utilise legacy data to find out new data points in the input space to close the coverage gap.
\end{enumerate}

Details of these two approaches will be discussed in the following subsections. 

\subsection{Frequency Domain Analysis Approach}
\label{subsec: freq domain}

    A frequency domain analysis, or AC analysis, is often used to extract useful information from a circuit. The time taken in carrying out an AC analysis is considerably less than doing a transient analysis which makes it more attractive to the analog design engineers to verify some fundamentals of the DUT and hence has been used some research works has well \cite{glsvlsi_2019, itc-india_2020}. 
    
    The AC small signal analysis first solves for the DC operating point values of the circuit. All non-linear devices in the circuit are abstracted as linear models utilising the DC operating point values. The AC node voltages of each node are computed as a function of frequency. A sinusoidal voltage source of small magnitude (say $1mV$) is placed at one of the input port ($n_I$). On selecting an output net ($n_O$) , the magnitude ratio and phase difference between the nets $n_I$ and $n_O$ is plotted as a function of frequency. Hence effectively, the AC analysis gives the Bode plot of the signal indicating the values of gain across different values of frequency of the input signal. With $\omega$ being frequency of the input signal, the Bode plot can be given as a function of $\omega$, denoted as $\mathcal{F}(\omega)$. 

    \subsubsection{AC Analysis and Range Coverage}

    We utilise this information to figure out what value operating frequency will be able to push the circuit to its extremes. The Bode plot is a good indication of how the amplitude of the signal at ouput port $n_O$ varies with the change of frequency of the signal applied at the input port $n_I$. Ideally, we should aim to operate the circuit near its pole to check if it goes into an unstable region or not. Poles are manifested as region in the Bode plot as local maxima. Hence, each local maxima is a point of interest to us. Let this frequency value be denoted as $\omega'$ such that, $\mathcal{F}(\omega') = max(\mathcal{F}(\omega))$. On keeping the amplitude of the input signal constant a higher gain will result in a higher amplitude at the output node. Thus applying an input having frequency $\omega'$ may lead to the increase in the amplitude of the output signal, that is, increasing the range coverage of it. The steps involved in the methodology are as follows.
    \begin{itemize}
        \item \textbf{Step 1}: Properly bias the circuit with its rated biasing voltages and currents. Additionally place an AC voltage source at the input terminal ($n_I$)
        \item \textbf{Step 2}: Perform an AC analysis and obtain the bode plot $\mathcal{F}(\omega)$ between nets $n_I$ and $n_O$.
        \item \textbf{Step 3}: Find the value of $\omega'$.
        \item \textbf{Step 4}: Simulate the circuit with an input signal having frequency $\omega'$.
    \end{itemize}
    
    An advantage of this approach is that it is time-efficient since AC analysis is much faster than traditional transient analysis. Another advantage lies in the fact that it gives a precise value of target frequency. Hence only one transient simulation is sufficient to drive the range of the output signal to higher or lower values. A disadvantage of this is that it is limited to range and deglitch coverage. 
    
    \subsubsection{Results on Benchmark Circuits}
    
    \begin{figure*}
		\centering
		\subfloat[Bode plot on AC analysis]{
			\includegraphics[width=0.45\textwidth]{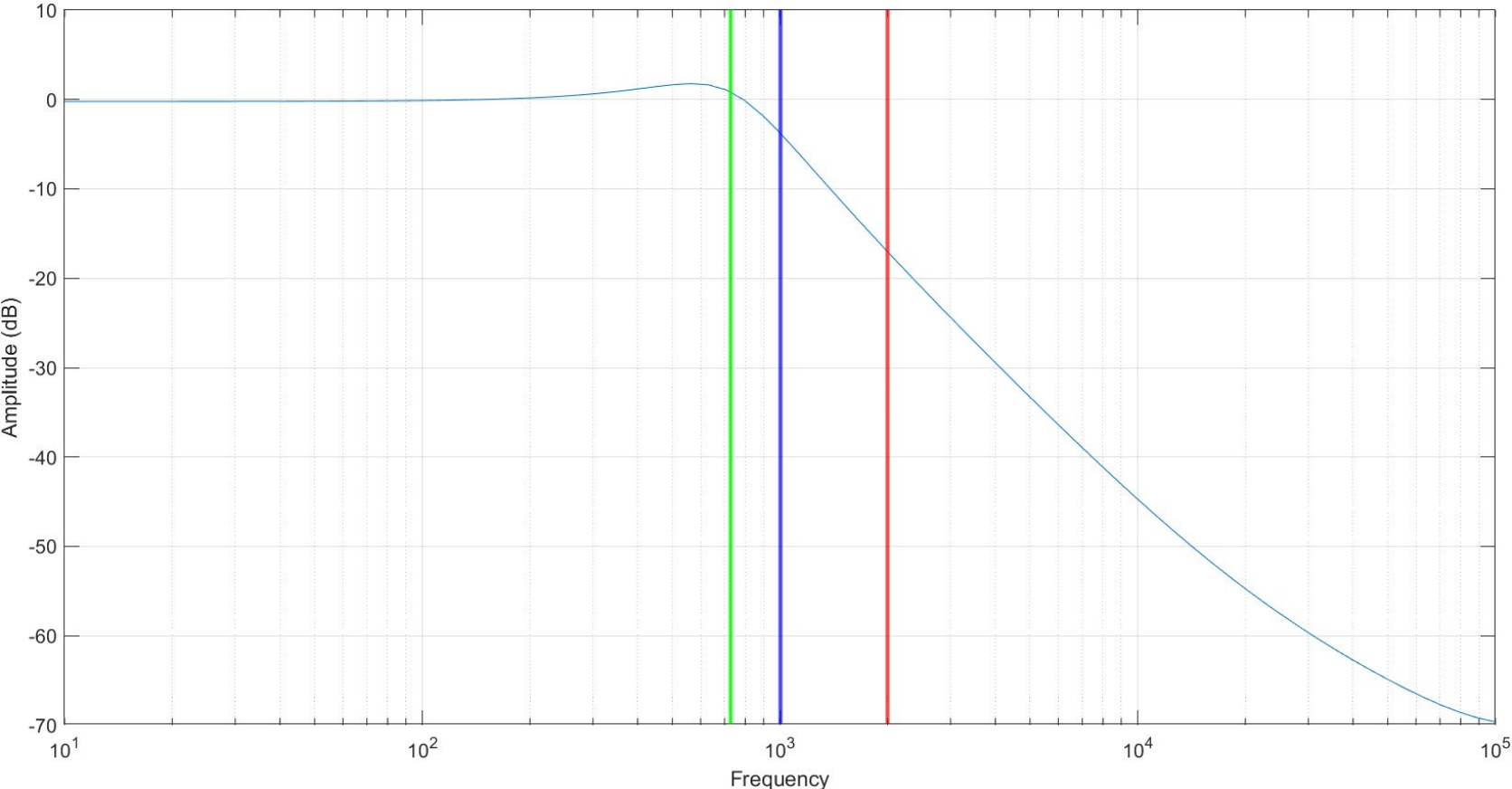}
			\label{fig: lpf bode}
			}\qquad
		\subfloat[Transient analysis at different input frequency]{
			\includegraphics[width=0.45\textwidth]{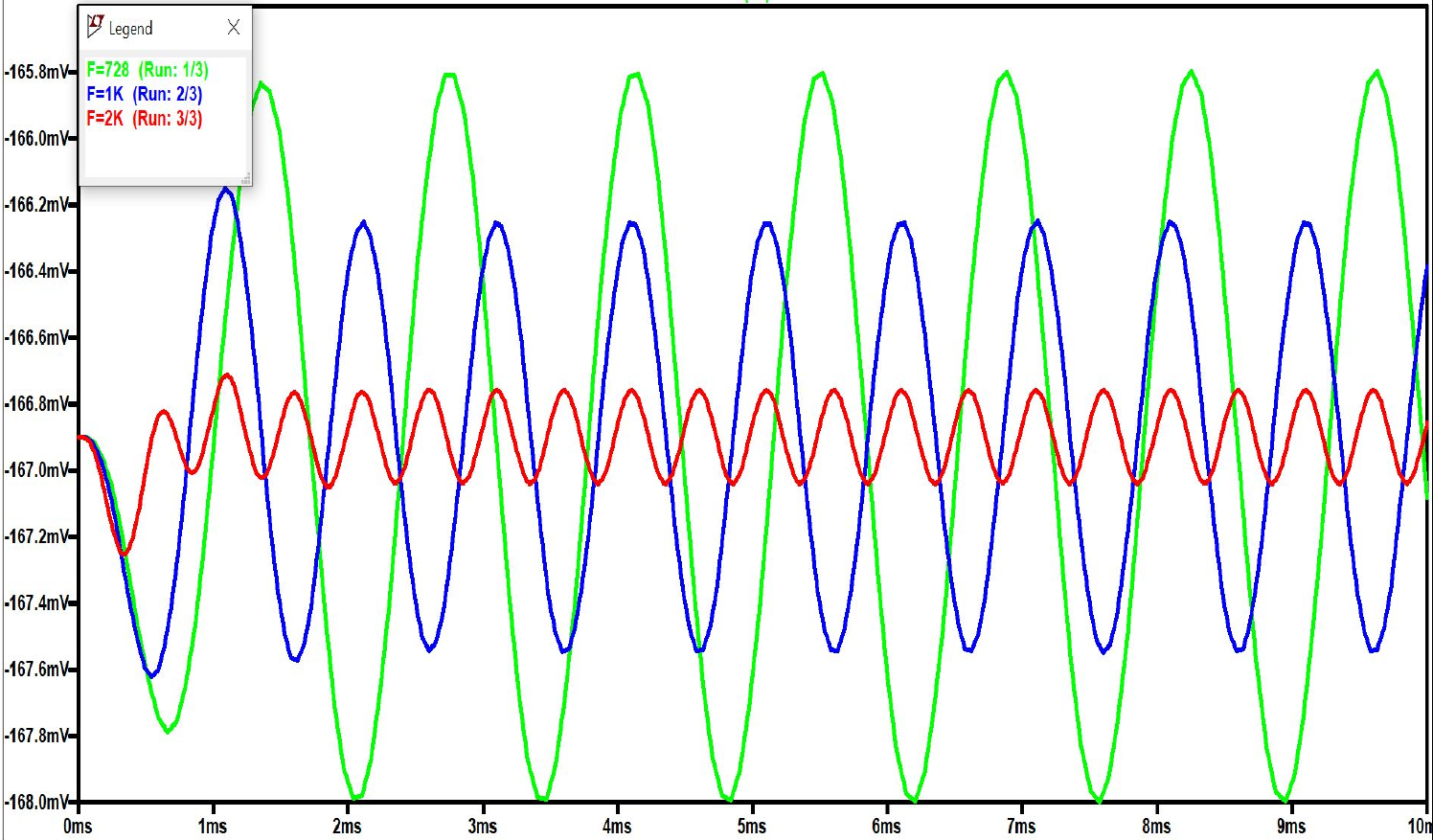}
			\label{fig: lpf tran}
		}
		\caption{AC and transient analysis of Low pass filter}
		\label{fig: lpf}
	\end{figure*}
	
	\begin{figure*}
		\centering
		\subfloat[Bode plot on AC analysis]{
			\includegraphics[width=0.45\textwidth]{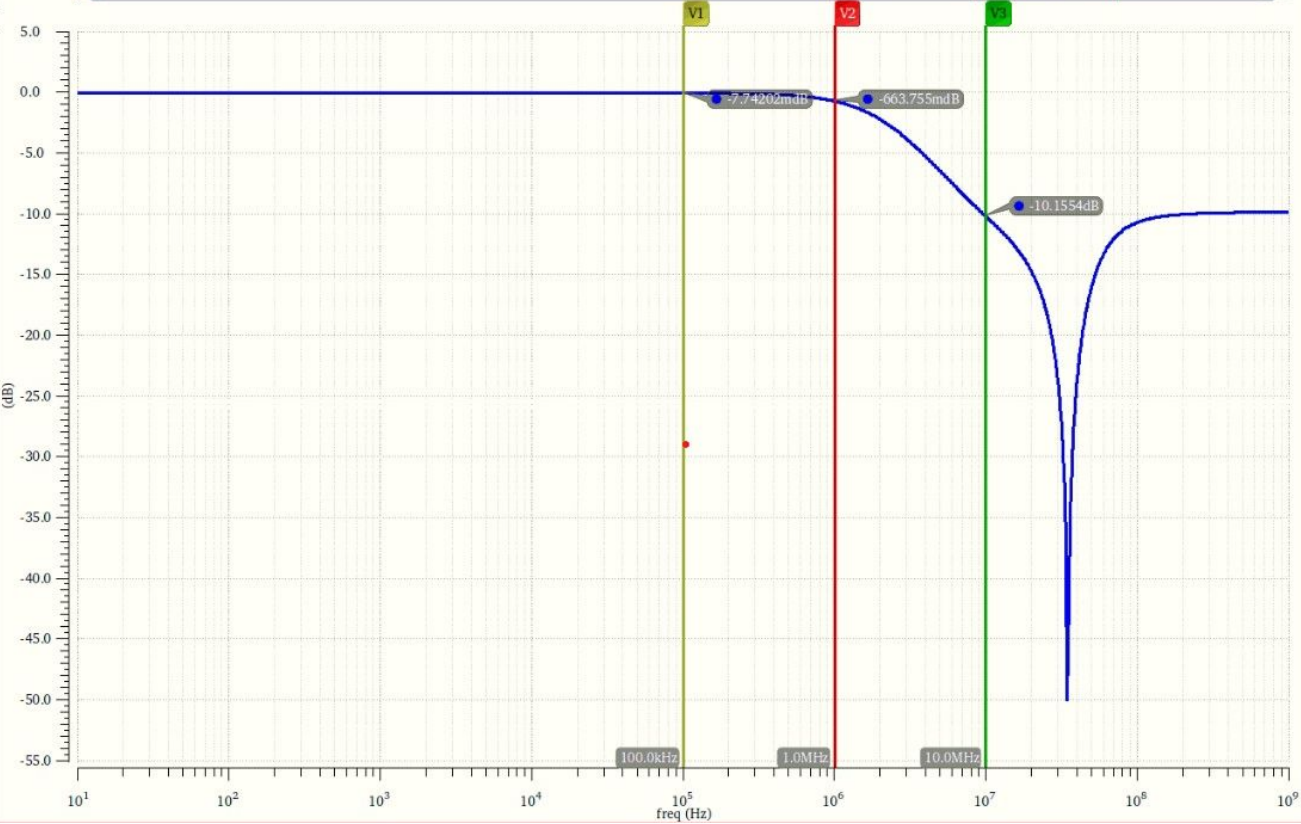}
			\label{fig: opamp1 bode}
		}\qquad
		\subfloat[Transient analysis at different input frequency]{
			\includegraphics[width=0.45\textwidth]{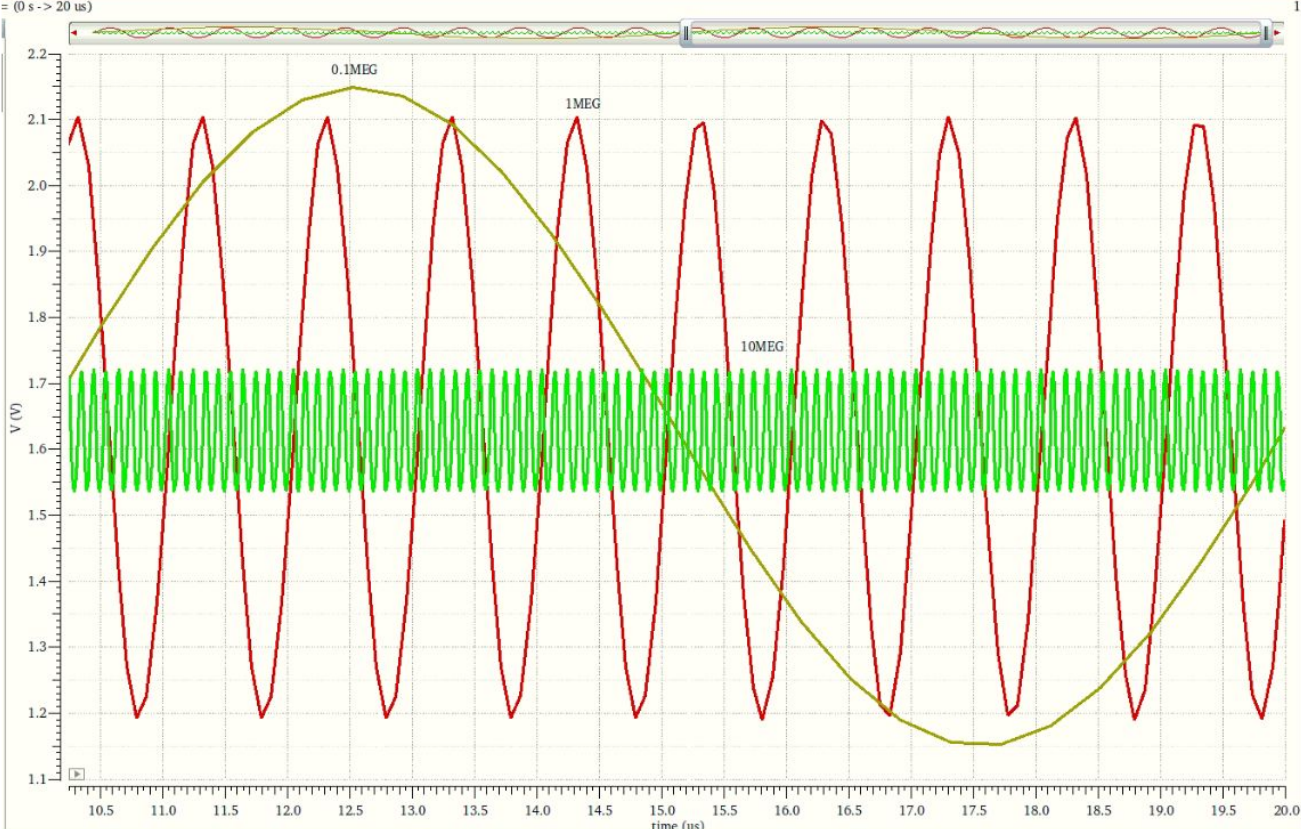}
			\label{fig: opamp1 tran}
		}
		\caption{AC and transient analysis of OPAMP1}
		\label{fig: opamp1}
	\end{figure*}
	
	\begin{figure*}
		\centering
		\subfloat[Bode plot on AC analysis]{
			\includegraphics[width=0.45\textwidth]{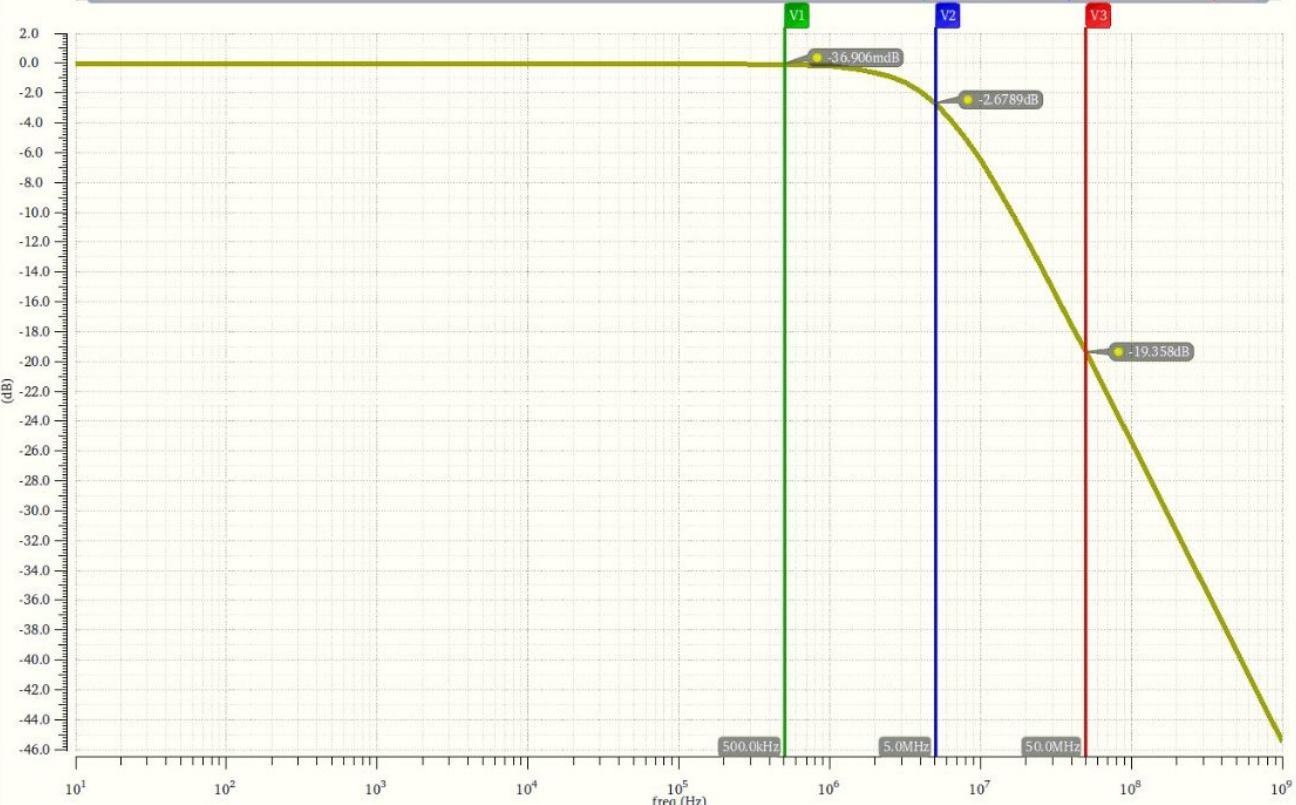}
			\label{fig: pll bode}
			}\qquad
		\subfloat[Transient analysis at different input frequency]{
			\includegraphics[width=0.45\textwidth]{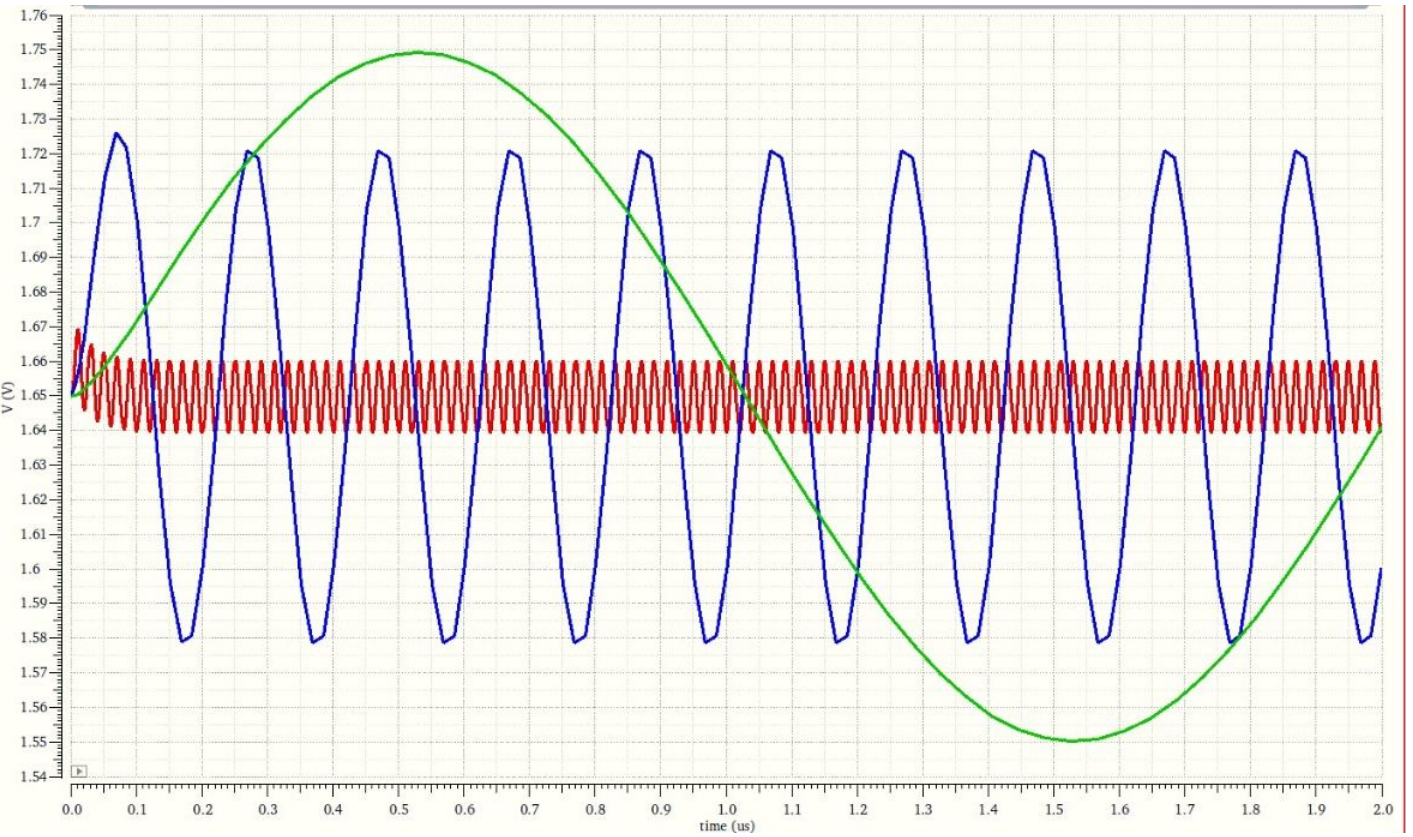}
			\label{fig: pll tran}
		}
		\caption{AC and transient analysis of PLL1}
		\label{fig: pll1}
	\end{figure*}
	
    We have tested our approach on two sets of AMS benchmark circuits, namely, ITC'97 \cite{itc_97} and ITC'17 \cite{itc_17}. Figure \ref{fig: lpf bode} shows the gain magnitude bode plot of the low-pass filter. Figure \ref{fig: lpf tran} shows the output waveform of the low-pass filter when subjected to transient analysis with input signals of varying frequencies. The frequencies of the input signal used during transient analysis are matched with the coloured vertical markers in the bode plot. As can be seen from the two figures, the Bode plot was attaining its maximum value at a frequency of $728 Hz$. The transient analysis with an input signal of the same frequency manifested the highest range of the output signal. The trend can also be seen at other two frequencies. Similar observations can be made from Figures \ref{fig: opamp1} and \ref{fig: pll1} which shows the results on OPAMP1 and PLL1 circuits from ITC'17 benchmark.

\subsection{Bayesian Optimization Approach}
\label{subsec: baysian optimization}

    Hyperparameter optimization \cite{hyperparameter_optimization} or black-box optimization strategies are often deployed when the function to be optimized is unknown and is thus treated as a black-box. Given a $k$-dimensional parameter space $\Omega \subseteq {\rm I\!R}^k $ it tries to search for an input point $x' \in \Omega$ such the following holds.
    $$x' = arg~ min_{x \in \Omega}~ f(x),$$
    where $y$ is the unknown black-box function. One obvious method is to query for the value of $f(x)$ for certain values of $x$ until we are sure that a minima has been reached. However there are certain restrictions for finding the value of $x'$. Firstly, we can not get the value of the gradient of the function at an input value $x$. Secondly each query is computationally expensive, therefore the number of queries should be kept as low as possible. One of the primary struggles with these optimizers is that they rely on manual tuning of the hyperparameter values. This requires manually searching through the input space and executing quite a few number of queries on the black-box function, which makes them computationally expensive. Though grid search and random search \cite{random_search} give slightly better performance, they lack the capability of inheriting the results of the past queries. 
    
    Bayesian optimization \cite{bayesian_optimization} is a type of black-box optimization problem where it tries to build an approximate estimation of the function, referred as {\em objective function} and hence tries to optimize it. It overcomes the previously mentioned limitation by keeping track of the results of previous queries. It is a sequential model-based approach. The model predicts a set of possible objective functions and a prior belief is prescribed over them. Let at any point during the iteration $\mathcal{D}$ denotes the available data and $\bm w$ is an unobserved random variable with a priori distribution $p(\bm w)$. $p(\bm w)$ denotes the {\em priori} beliefs about the probable values that $\bm w$ can attain. With this in hand, Bayes rule can be used to compute the {\em posteriori} distribution $p({\bm w}| \mathcal{D})$ as follows. 
    \[
        p({\bm w}| \mathcal{D}) = \frac{p(\mathcal{D} | {\bm w})p(\bm w)}{p(\mathcal{D})}
    \]
    The above {\em posterior} probability distribution describes potential values for $f(x)$ at a candidate point $x$. As can be inferred from above, each time a new data point is brought into consideration, the posterior probability distribution gets updated. An {\em acquisition} function is also used to measure the value that a new candidate point $x$ will give based on the current {\em posterior} distribution. The {\em acquisition} function can be tuned according to the application requirements. 
    
    In here we briefly describe {\em expected improvement} acquisition function which is the most commonly used one. Suppose after $n$ iterations $f^*_n = min_{m\leq n}f(x_m)$ is the minimum value that has been observed. Ideally, we should place the next query on a point $x$ such that we are confident about getting a lower value of $f(x)$ than $f^*_n$. Thus the improvement in the value that will be observed can be given as $(f^*_n - f(x))$, if $f(x) < f^*_n$, else $0$. The expression can be re-written as $[f^*_n - f(x)]^+$, where $[p]^+ = max(p, 0)$. Since $f(x)$ is unknown to us, we take the expected value of this improvement term and take the value of $x$ which maximises it. The {\em expected improvement}, EI, is defined as
    \[
        EI_n(x) := E_n[[f^*_n - f(x)]^+],
    \]
    where $E_n[\cdot] = E[\cdot | x_1, x_2, \cdots, x_n, f(x_1), f(x_2), \cdots , f(x_n)]$ denote the expectation taken under the {\em posterior} distribution for the valuations of $f$ at $x_1, ~x_2,~ \cdots, x_n$. A closed form solution of this expression has been given by \cite{jones_1998} as follows.
    
    \[
        EI_n(x) = [\Delta_n(x)]^+ + \sigma_n(x)~ \varphi \left ( \frac{\Delta_n(x)}{\sigma_n(x)} \right ) - |\Delta_n(x)| \Phi \left ( \frac{\Delta_n(x)}{\sigma_n(x)} \right ),
    \]
    where $\Delta_n(x) := f^*_n - \mu_n(x)$ is the expected quality improvement by bringing in new candidate point $x$. Finally the next candidate point $x_{n+1}$ is choosen as the one which gives the maximum expected improvement, 
    \[
        x_{n+1} = arg~ max~ EI_n(x)
    \]
    
\begin{figure*}
		\centering
		\subfloat[Objective function]{
			\includegraphics[width=0.45\textwidth]{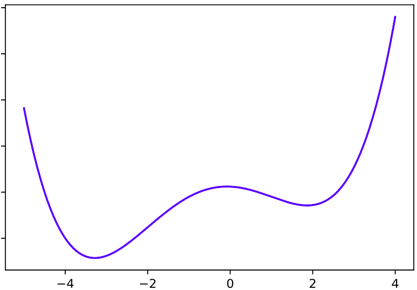}
			\label{fig: objective function}
			}\qquad
		\subfloat[Learned function]{
			\includegraphics[width=0.45\textwidth]{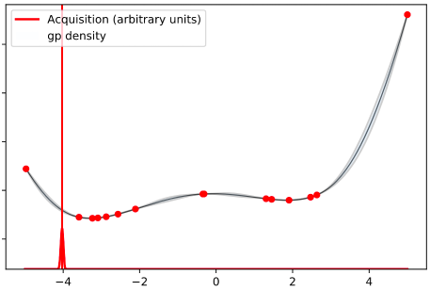}
			\label{fig: learned function}
		}
		\caption{An example of Bayesian optimization}
		\label{fig: bayesian optimization}
	\end{figure*}

    Figure \ref{fig: bayesian optimization} shows a demonstration of how Bayesian optimization works on a univariate function. Figure \ref{fig: objective function} shows the objective function required to optimize. Figure \ref{fig: learned function} shows the learned function that the Bayesian optimization algorithm is able to produce. The red dots on the learned function are the input points for which the optimizer had placed a query. As can be seen from the figure, the next query will be placed near the value $-4$ where the current formulation of the probability density function attains the highest value.
    
    Bayesian optimization techniques have been applied to analog circuit domain in the past. In \cite{peng_li_iccad18, peng_li_dac19} researchers have utilised and tuned Bayesian optimization technique to search for rare fault scenario in AMS designs. The channel length, threshold voltage, and gate oxide thickness were chosen as the input parameters. A fault scenario is modelled as the one where the circuit operation goes out of the thresholds set by the specification.

	\begin{figure*}
		\centering
		\subfloat[Extending the range of output voltage of LDO on controlling the load current]{
			\includegraphics[width=0.45\textwidth]{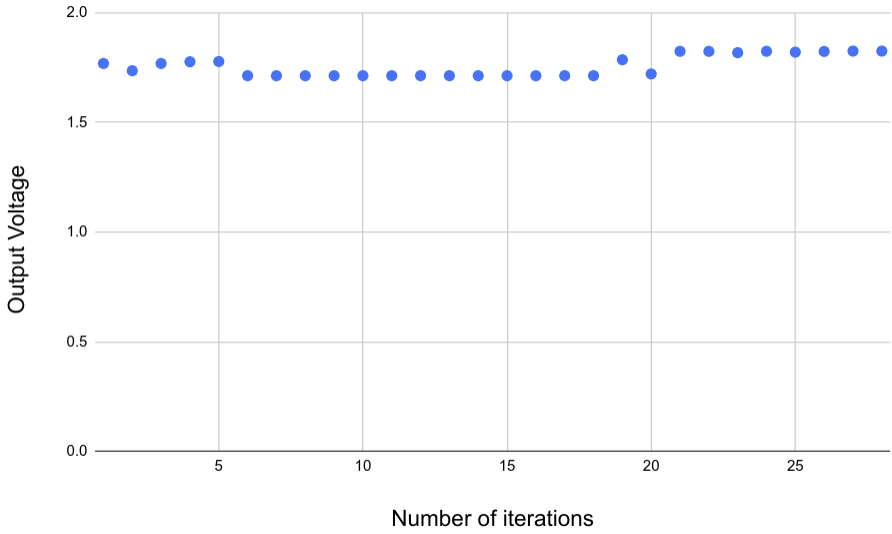}
			\label{fig: ti ldo regression}
		}\qquad
		\subfloat[Objective function as approximated by optimizer. X-axis denotes load current, Y-axis represents output voltage level]{
			\includegraphics[width=0.45\textwidth]{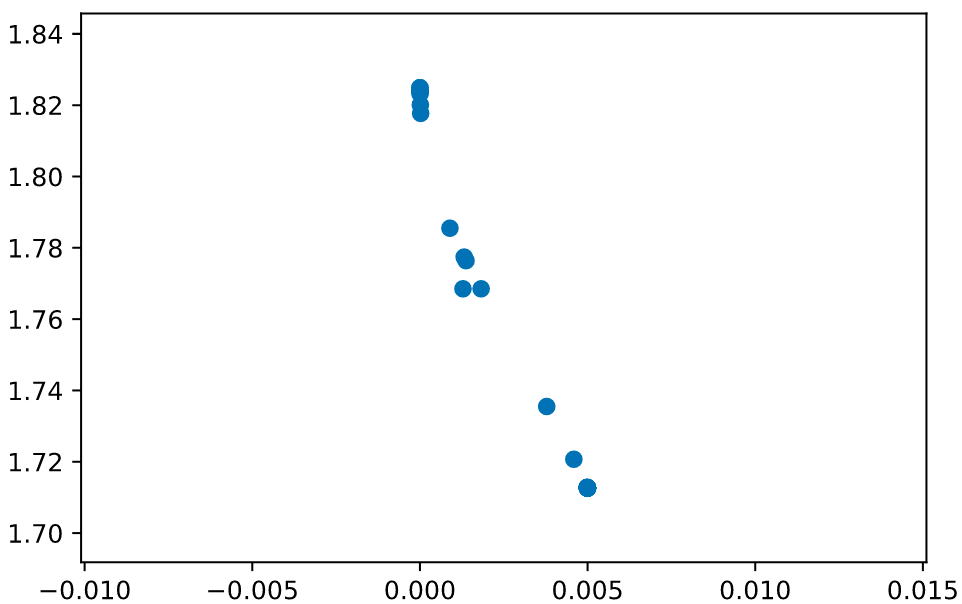}
			\label{fig: ti ldo function}
		}
		\caption{Applying Bayesian optimisation on to find corner cases in an industrial LDO by TI}
		\label{fig: ti ldo}
	\end{figure*}
	
	\begin{figure*}
		\centering
		\subfloat[Extending the range output signal frequency of OSC1 on varying the supply voltage]{
			\includegraphics[width=0.45\textwidth]{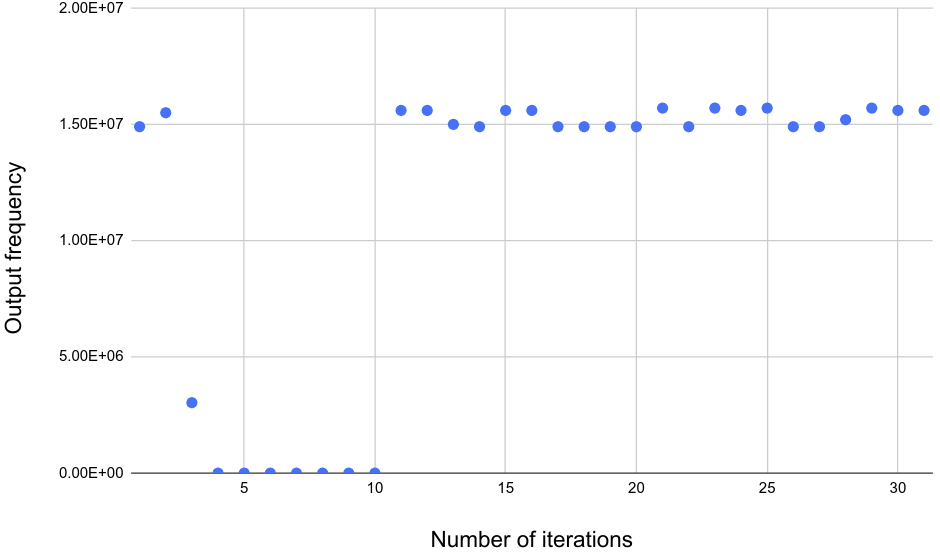}
			\label{fig: osc1 regression}
		}\qquad
		\subfloat[Objective function as approximated by optimizer. X-axis denotes value of supply voltage, Y-axis represents output signal frequency]{
			\includegraphics[width=0.45\textwidth]{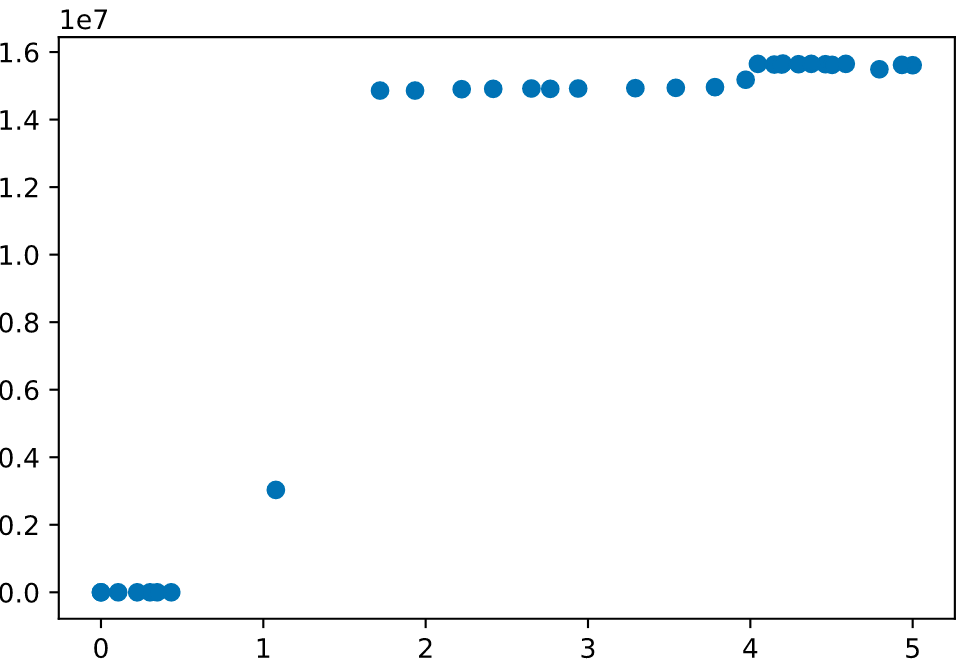}
			\label{fig: osc1 function}
		}
		\caption{Applying Bayesian optimisation on to find corner cases in OSC1 from ITC'17 AMS benchmark circuit}
		\label{fig: osc1}
	\end{figure*}

    \subsubsection{Bayesian optimizer for analog coverage}
    In the application domain of analog coverage, the input space $I_{valid}$ is the input parameter space while $f(I_{valid})$ constitutes the output space. The DUT ($f$) serves as the black-box. It is to be noted, that the function $f$ is only for representation purpose and will not be the same while computing output values for all coverpoints. The output of a range coverage artefact will obviously vary from that of a frequency coverage artefact. Hence on application of the Bayesian optimization technique different functions have to be estimated depending on the coverpoints that they correspond to. In the context of analog coverage, they are primarily two areas where Bayesian optimization can be applied. They are listed as follows.
    
\begin{enumerate}
    \item {\bf Decreasing the coverage gap}: As mentioned earlier, coverage gap is the region in the target specification that has not been covered (visited during simulation) yet. Bayesian optimization can be used to guide the simulation towards these values by efficiently choosing the right set of input parameter values. For a given coverpoint $C$, and a target interval $[a:b]$ for $C$ we try to search for two valuations in the input parameter space ,$x_{min}$ and $x_{max}$, such that the following satisfies
    $$y_C(x_{min}) = min(y_C(x) - a)$$
    $$y_C(x_{max}) = min(b - y_C(x))$$
    
    where $y_C$ is the black-box objective function for mapping input parameter value points to output space for coverpoint C.
    
    \item {\bf Bug detection}: As seen from Figure \ref{fig: coverage overview}, a portion of $f(I_{valid})$ may also invades into the {\em forbidden states}. In coverage specification terms, the {\em forbidden state} is represented as {\em illegal bins}. A visit to any of these illegal bins is an indication of a probable bug in the design and should be flagged. Most of the easy-to-hunt bugs get manifested in an earlier stage of the design thus trimming the portion of $f(I_{covered})$ which has an intersection with the {\em forbidden states}. Practically, at a mature stage of the design cycle when all the necessary guards are in place, finding an input combination which can lead the DUT to such a region is a non-trival task. Extensive amount of simulations are carried out to ensure that all such bugs are caught. However, no such abstraction exists that can quantify this continuous state space. Thus even after extensive simulations, one can not guarantee that the $f(I_{covered})$ is devoid of any forbidden state. We apply Bayesian regression in this targeted context by trying to choose the input parameter values in such a way that it visits one of the illegal bins. Hence for a given coverpoint $C$ and an illegal bin interval as $[c:d]$, the Bayesian optimization problem can be given as follows. 
    
    $$min_{x\in \Omega}~  \big | y_C(x) - \frac{d-c}{2} \big |$$
    
    where $y_C$ is the black-box objective function for mapping input parameter value points to output space for coverpoint C.
    
\end{enumerate}

\subsubsection{Results}

    The above mentioned approach was applied on two circuits. This subsection presents the results on both the circuit. Our first case study was an industrial LDO developed by TI. The input space consisted of the load current the LDO was driving while the value of output voltage level of the LDO is our output entity. The domain of $I_{valid}$, that is, the permissible range of load current was set as $[0:0.5]$. Figure \ref{fig: ti ldo regression} shows how the Bayesian optimizer searched for the extreme values as the number of iteration progressed. The approximated objective function is shown in Figure \ref{fig: ti ldo function}. Within the supplied domain of $I_{valid}$ the maximum output voltage level of $1.8249$ was achieved for a load current of $1\mu A$, while minimum of $1.7126 V$ was reported for $0.00499947 A$.

    The next test case constituted of {\em OSC1} circuit from the ITC'17 AMS benchmark circuits \cite{itc_17}. The value of supply voltage was chosen as the independent input variable which got manifested as the frequency of the output signal as was hence captured by a frequency coverage artefact. Figure \ref{fig: osc1} shows the Bayesian regression and objective function in a similar fashion.

\section{Conclusion}
\label{sec: conclude}
	Finding an input vector that can drive an analog circuit into a specific narrow region is a needle in a haystack problem. However, some heuristics and previous knowledge about the system can aid us in pruning down the search space and accelerate the searching task. We have shown two such techniques which are able to find a specific input vector for an analog system, such that its response is mapped into a novel region. This region may act in closing down the coverage gap in the verification strategy or may activate a hidden bug in the system such that it can be get fixed. We have also demonstrated the working of our methodology on some AMS benchmark circuits and an industrial circuit.

\balance

\end{document}